# Microwave and RF Applications for Micro-resonator based Frequency Combs

Thach G. Nguyen,[1] Mehrdad Shoeiby,[1] Marcello Ferrera,[2] Alessia Pasquazi,[3]
Marco Peccianti,[3] Sai T. Chu,[4] Brent E. Little,[5] Roberto Morandotti,[6] Arnan Mitchell,[1]
and David J. Moss [7]

[1]Schoof of Electrical and Computer Engineering, RMIT University, Melbourne, VIC 3001, Australia
[2]School of Engineering and Physical Sciences, Heriot-Watt University,
David Brewster Building, Edinburgh, Scotland EH14 4AS, United Kingdom
[3]Department of Physics and Astronomy, University of Sussex, Falmer, Brighton BN1 9RH, United Kingdom
[4]Department of Physics and Materials Science, City University of Hong Kong, Tat Chee Avenue, Hong Kong, China
[5]Xi'an Institute of Optics and Precision Mechanics of CAS, Xi'an, China
[6]INSR – Énergie,Matériaux et Télécommunications, 1650 Blvd Lionel Boulet, Varennes (Québec), J3X1S2, Canada
[7]Center for Micro-Photonics, Swinburne University of Technology, Melbourne, VIC Australia
dmoss@swin.edu.au

*Abstract*

Photonic integrated circuits that exploit nonlinear optics in order to generate and process signals all-optically have achieved performance far superior to that possible electronically - particularly with respect to speed. We review the recent achievements based in new CMOS-compatible platforms that are better suited than SOI for nonlinear optics, focusing on radio frequency (RF) and microwave based applications that exploit micro-resonator based frequency combs. We highlight their potential as well as the challenges to achieving practical solutions for many key applications. These material systems have opened up many new capabilities such as on-chip optical frequency comb generation and ultrafast optical pulse generation and measurement. We review recent work on a photonic RF Hilbert transformer for broadband microwave in-phase and quadrature-phase generation based on an integrated frequency optical comb. The comb is generated using a nonlinear microring resonator based on a CMOS compatible, high-index contrast, doped-silica glass platform. The high quality and large frequency spacing of the comb enables filters with up to 20 taps, allowing us to demonstrate a quadrature filter with more than a 5-octave (3 dB) bandwidth and an almost uniform phase response.

## I. Introduction

Photonic integrated circuits that exploit nonlinear optics in order to achieve all-optical signal processing have been demonstrated in silicon-on-insulator (SOI) nanowires [1] and chalcogenide glass (ChG) waveguides [2, 3]. Some of the key functions that have been demonstration include all-optical logic [4], time division de-multiplexing from 160Gb/s [5] to over 1Tb/s [6], to optical performance monitoring using slow light at speeds of up to 640Gb/s [7-9], all-optical regeneration [10,11], and many others. The 3$^{rd}$ order nonlinear efficiency of all-optical devices depends on the waveguide nonlinear parameter, $\gamma = \omega\, n_2 / c\, A_{eff}$ (where $A_{eff}$ is the waveguide effective area, $n_2$ the Kerr nonlinearity, and $\omega$ the pump frequency). It can also be increased by using resonant structures to enhance the local field intensity. High index materials, such as semiconductors and ChG, offer excellent optical confinement and high values of $n_2$, a powerful combination that has produced extremely high nonlinear parameters of $\gamma$= 200,000 W$^{-1}$ km$^{-1}$ for SOI nanowires [1], and 93,400 W$^{-1}$ km$^{-1}$ in ChG nanotapers [2].

However, silicon suffers from high nonlinear losses due to intrinsic two-photon absorption (TPA) [1] and the resulting generated free carriers that tend to have very long lifetimes. Even if free carriers are eliminated by using p-i-n junctions, however, silicon's very poor intrinsic nonlinear figure of merit (FOM = $n_2$ / ($\beta\, \lambda$), where $\beta$ is the two-photon absorption coefficient) of around 0.3 in the telecom band is too low to achieve extremely high performance all-optical processing. The consequences of silicon's low FOM was clearly illustrated in breakthroughs at longer wavelengths in the SWIR band, where the TPA vanishes [12-14]. While TPA can sometimes be used to advantage for all-optical functions [15-17], for the most part in the telecom band silicon's low FOM poses a fundamental limitation, since it is a material property that is intrinsically dependent on the bandstructure and so it cannot be improved.

In 2008-2010, new platforms for nonlinear optics that are CMOS compatible were introduced, including Hydex and silicon nitride [18-34]. These platforms exhibit negligible nonlinear absorption in the telecom band, and have since been the basis of many key demonstrations. They have revolutionized micro-resonator optical frequency combs [20] as well as ultrashort modelocked lasers [22]. They show great promise for applications to ultrahigh bandwidth telecommunications [34]. Furthermore, they have also shown promising performance for applications to quantum optics [35, 36].

The first integrated CMOS compatible integrated optical parametric oscillators were reported in 2010 [20, 33], showing that Kerr frequency comb sources could be realised in chip form by using ring resonators with relatively modest Q-factors, in comparison with the extremely high Q's of micro-toroid structures [37]. Continuous wave "hyper-parametric" oscillation in a micro-ring resonator with a Q factor of 1.2 million was demonstrated with a differential slope efficiency 7.4% for a single oscillating mode out of a single port, a low CW threshold power of 50mW, and a variable range of frequency spacings from 200GHz to > 6THz. Following this, a stable modelocked laser with pulse repetition rates from 200GHz to 800GHz [22] was demonstrated.

The success of this platform arises from its very low linear loss as well as moderately high nonlinearity parameter ($\gamma \cong 233 W^{-1} km^{-1}$) and, most importantly, negligible nonlinear loss (TPA) up to intensities of 25GW/cm$^2$ [21]. The low loss, design flexibility, and CMOS compatibility of these devices will go a long way to achieving many key applications such as multiple wavelength sources for telecommunications, computing, metrology and many other areas.

Many applications, including radar mapping, measurement, imaging as well as the realization of advanced modulation formats for digital communications, require the generation, analysis and processing of analogue RF signals where both the amplitude and phase of the signals are critically important. In order to access such information, it is often necessary to perform a uniform quadrature-phase shift (±90°) of the constituent RF frequencies over the entire bandwidth of interest. The process of obtaining in-phase and quadrature-phase components of a signal can be achieved via a Hilbert transform, and for RF signals this is often achieved using 'hybrid coupler' [38] technology that, while mature, is often degraded by large amplitude and phase ripple [39 - 41]. In addition, it is difficult to achieve wide band operation using electronic circuits. In many applications, especially those involving radar and early warning receivers in electronic warfare systems, the capacity to process signals over a multi-octave spectral range from below 1 GHz to 20 or even 40 GHz is generally required [42]. Electronic approaches to microwave signal processing often need banks of parallel systems to cover such a wide spectral bandwidth.

Photonic implementations of Hilbert transformers have been demonstrated that have achieved very high performance over a very wide bandwidth. Furthermore, these devices are immune to electromagnetic interference, while being compatible with fiber remote distribution systems and parallel processing. There are several approaches to implementing photonic Hilbert transformers, including phase-shifted Bragg grating devices [43-45] or Fourier domain optical processors [46]. However, the performance in terms of bandwidth and low frequency cutoff typically associated with Hilbert transformers based on phase-shifted Bragg gratings has been limited due to the difficulty of fabricating grating devices with the necessary stringent requirements. Programmable Fourier domain optical processors [46] have been proposed as an approach to greatly reduce amplitude and phase ripple over very broad RF bandwidths; however, even this approach suffers from performance degradation at frequencies below 10 GHz due to typical resolution limits of Fourier domain optical processors. Recently, a tunable fractional temporal Hilbert transformer based on the phase shift at the resonant wavelength of a ring resonator has been demonstrated [47]. Although the Hilbert transformers based on grating or ring resonator structures can provide a phase shift at the center frequency within the device bandwidth, accessing both the in-phase and quadrature phase components of the same signal is not available.

Transversal filtering is a versatile approach that can achieve a diverse range of signal processing functions. In this approach, the signal is divided into a number of copies (taps) with different weights, each copy having a tap that is delayed in time by incremental amounts. All of the delayed taps are then summed and detected. Each of the taps can be considered a discrete sample of the filter impulse response. By carefully controlling the amplitudes and delays of all taps, different filtering functions can be realized.

Microwave Photonic transversal filters have been reported by several groups [48], including a broadband photonic Hilbert transformer for in-phase and quadrature-phase generation [49, 50] that exhibited low amplitude ripple and very low phase error over a multi-octave bandwidth. The application of this transformer to instantaneous frequency measurement (IFM) has been reported [51-53]. Photonic transversal filtering has typically relied on multiple discrete laser diodes for the multi-wavelength source. However, since the filter bandwidth strongly depends on the number of filter taps (i.e. the number of wavelengths) and therefore the number of laser diodes, this typically results in a very high cost, complexity, energy consumption, and footprint, particularly for systems that are parallelized. Therefore, as an alternative to using individual lasers for each tap, a single component that can generate multiple, and high quality, wavelengths would be highly advantageous.

High quality, large spectral range multi-wavelength sources have typically been based on mode-locked fiber lasers [54], electro-optically generated combs [55] and nonlinear micro-resonators [56]. Of these, integrated optical frequency combs based on the Kerr nonlinearity in nonlinear micro-resonators have displayed significant potential for high performance transversal filtering for RF signal processing. These devices greatly reduce both cost and complexity due to their very compact size as well their suitability for monolithic integration. As discussed above, significant progress has been made in demonstrating high quality, wide spectrum optical frequency combs [20, 22, 25, 27, 33, 57, 58]. Moreover, unlike the combs generated from mode-locked lasers or RF modulation, micro-resonators can achieve very large frequency spacings [27], allowing for the control of the amplitude and phase of individual comb lines via commercially available wave shapers [59]. Indeed, comb sources based on silicon nitride microrings have been successfully used for high performance programmable band-pass RF filter [59]. This application of micro-resonator frequency combs to RF filtering has added an important dimension to ultrahigh bandwidth RF and microwave applications of nonlinear photonics [60-66].

In this paper, we review recent work on a photonic Hilbert transformer [60] based on multi-tap transversal filtering. We use an integrated frequency comb source based on a high Q factor nonlinear microring resonator fabricated in a CMOS compatible, high-index, doped-silica glass (Hydex) waveguide platform [20, 27]. The comb source has a wide spectral range with a large frequency spacing of 200 GHz, allowing for the realization of filters with up to 20 taps. We demonstrate an RF quadrature coupler with more than a 5-octave 3dB bandwidth and with a near constant relative phase over the pass band. This represents the first demonstration of a wide band photonic RF Hilbert transformer, or indeed any transversal filter, based on an integrated comb source that has both positive and negative taps.

## II. INTEGRATED COMB SOURCE

Figure 1 shows a ring resonator that was the basis of the integrated optical parametric oscillator. It is a four port micro-ring resonator with radius $\cong$ 135μm in waveguides havng a cross section of 1.45μm x 1.5μm, buried in $SiO_2$. The waveguide core is Hydex glass with n=1.7 and a core-cladding contrast of 17% [20]. The glass films were deposited by plasma enhanced chemical vapor deposition (PECVD) and subsequently processed by deep UV photolithography using stepper mask aligners followed by dry reactive ion etching, before over-coating with silica glass. Propagation losses were < 0.06dB/cm and pigtail coupling losses to standard fiber was $\cong$ 1.5dB / facet. The ring resonator had a FSR of 200GHz and a resonance linewidth of 1.3pm, with a Q of 1.2 million. The dispersion is anomalous [18] over most of the C-band with a zero dispersion point for TM polarization near ~1560nm with λ < 1560nm being anomalous and λ > 1560nm normal.

Figure 1 shows the OPO spectrum for a TM polarized pump at 1544.15nm at a power of 101mW. Oscillation initiates near 1596.98nm, 52.83nm away from the pump, with frequency spacing of 53nm. This agrees well with the calculated peak in the modulational instability (MI) gain curve which occurs near ~1590nm. Figure 1 also shows the output power of the mode at 1596.98nm from the drop port, versus pump power, showing a differential slope efficiency of 7.4%. The maximum output power at 101mW pump at 1544.15nm was 9mW in all modes out of both ports, representing a total conversion efficiency of 9%. When pumping at 1565.19 nm in the normal dispersion regime, oscillation was not achieved. Pumping at zero dispersion (1558.65nm) resulted in a spacing of 28.15nm, which also agreed with the peak in the MI gain profile.

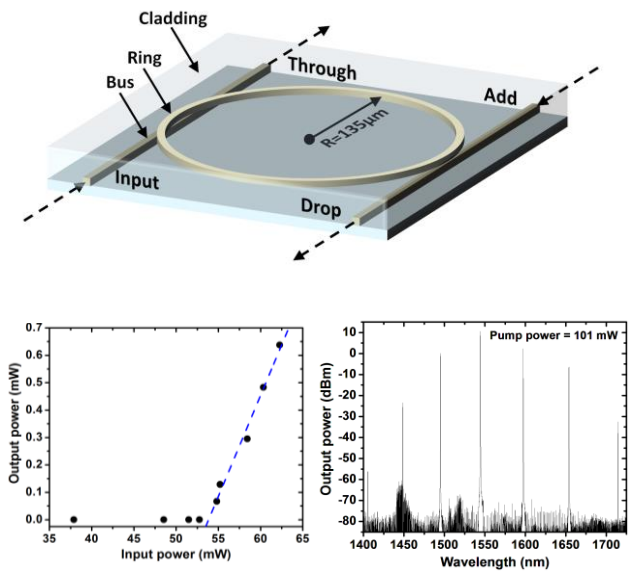

Figure 1. Top: Hydex based micro-ring resonator with Q factor of 1.2M. Bottom: Spectra of hyper-parametric oscillator at full pumping power (101mW, right). Output power of a single line, single port vs pump power (bottom, left).

## III. PHOTONIC HILBERT TRANSFORMER VIA TRANSVERSAL FILTERING

An ideal Hilbert transformer exhibits a constant amplitude frequency response and a ±90° frequency independent phase shift centred around the main (central) frequency. The impulse response of an ideal Hilbert transformer [Fig. 2(a)] is a continuous hyperbolic function $1/(\pi t)$ that extends to infinity in time. In order to realize this impulse response using transversal filtering, the hyperbolic function is truncated and sampled in time by discrete taps [49]. The theoretical RF transfer function of the filter is the Fourier transform of the impulse response. Fig. 2 (c) shows the calculated spectrum associated with the filtered signal amplitude when the impulse response is truncated and evenly sampled. Here, the bandwidth is limited by amplitude ripple, with 'nulls' at frequencies zero and $f_c$. The null frequency $f_c$ is determined by the sample spacing $\Delta t$ as $f_c = 1/\Delta t$. Amplitude ripples and the bandwidth in particular depend strongly on the number of taps, where the bandwidth increases dramatically with the number of sample taps. For example, the 3 dB bandwidth increases from less than 3 octaves with a 4 tap filter to more than 5 octaves when the number of filter taps is increased to 16 or more.

Figure 2(d) shows the calculated frequency response of the filtered signal phase for different numbers of taps in the impulse response. Unlike the amplitude frequency response, when the impulse response function is truncated and sampled, the phase response does not exhibit any ripples from zero up to the null frequency $f_c$. The phase is constant at -90° regardless the number of samples. At the null frequency, $f_c$, the phase transitions from -90° to 90°.

The simulations presented in Fig. 2 indicate that a band-limited Hilbert transformer can be realized using a transversal filtering method with the tap coefficients set to a hyperbolic function. The bandwidth and pass-band amplitude ripples are determined by the spacing between filter taps and by their number. The Hilbert transform impulse response is an asymmetric function centred at time zero. This has two implications for realizing a practical device – first, a reference for time 'zero' is required, as illustrated in Fig. 2(a). Secondly, negative tap coefficients are required. Practical implementations of photonic transversal filters are often realized by assigning filter taps to different optical wavelengths. The time delay between filter taps (wavelengths) can be achieved by using a dispersive medium such as a long optical fiber [49] or a chirped grating [52]. In previous demonstrations of a Hilbert transformer using photonic transversal filtering, discrete filter taps were realized by employing an array of discrete continuous-wave laser sources. However, this approach limits the number of filter taps to only four, resulting in less than a 3-octave bandwidth. In this work, we use a wide spectrum integrated comb source based on a high-Q microring resonator, thus allowing a much higher number of filter taps, in turn significantly broadening the RF bandwidth.

Figure 3 illustrates the experiment setup, which consists of three main sections: frequency comb generation, comb shaping, and photonic RF filtering. A continuous-wave (CW) tunable laser amplified by a high power Erbium-doped fiber amplifier (EDFA1), was used as the pump source for the microring resonator. The equally spaced comb lines produced by the ring resonator were then shaped according to the required tap coefficients using a reconfigurable filter, having a much smaller resolution than the comb line spacing. The waveshaper also split the comb into two paths, which were connected to the photonic RF filter section. Comb lines corresponding to positive tap coefficients were routed to one of the output ports of the waveshaper, while comb lines corresponding to negative tap coefficients were sent to the second port.

The two outputs of the waveshaper were connected to two inputs of a 2x2 balanced Mach-Zehnder modulator (MZM), biased at quadrature, where the comb lines were modulated by the RF signal. One group of comb lines was modulated on the positive slope of the MZM transfer function while the second group was modulated on the negative slope. This allowed both negative and positive tap coefficients to be realized with a single MZM. The output of the MZM was then passed through 2.122 km of single-mode fiber that acted as a dispersive element to delay the different filter taps. The dispersed signal was then amplified by a second fiber amplifier (EDFA2) to compensate for loss, and filtered in order to separate the comb from the signal at the pump wavelength in order to produce the system 0° phase reference. The second path was used as the 90° phase signal. The signal path was time-shifted with a variable optical length (VOL) so that the reference could be positioned, in time, exactly at the middle of the filter taps, as illustrated in Fig. 2(a).

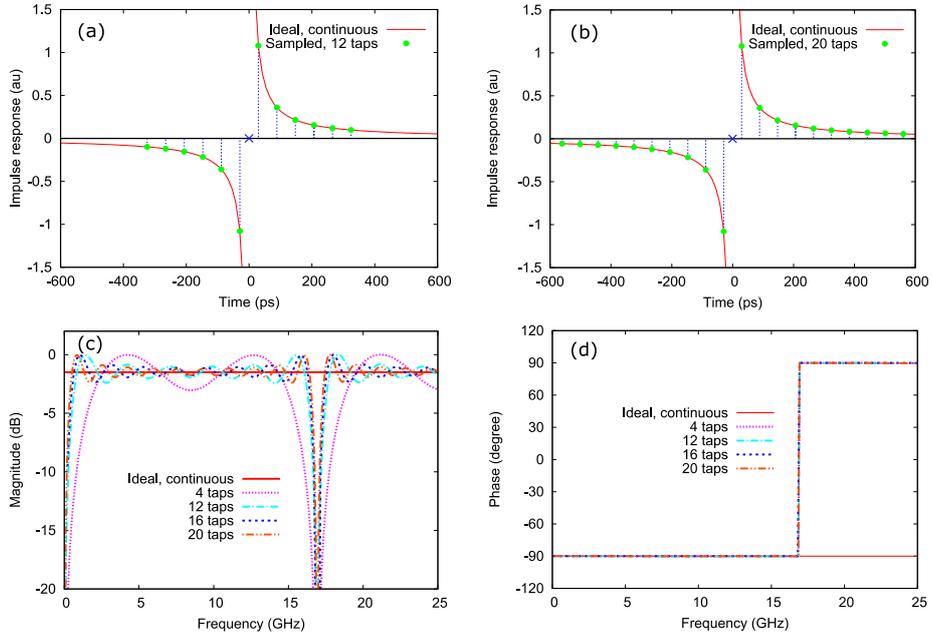

Fig. 2. Working principle of a transversal Hilbert transformer: (a) - (b) Ideal, continuous hyperbolic and discretely sampled impulse response with time spacing Δt = 59 ps. Corresponding (c) amplitude and (d) phase response for the continuous and discrete cases, respectively.

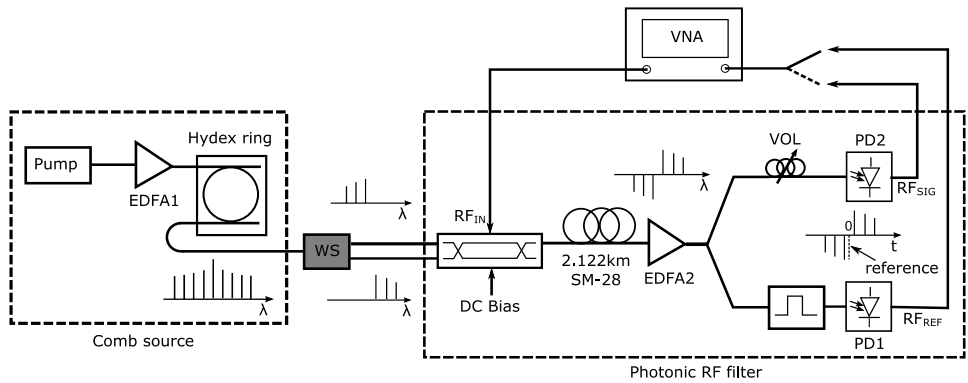

Fig. 3. System implementation of a Hilbert transformer exploiting a microring-based comb source. VNA is a Vector Network Analyzer, WS= wavelength selective switch, VOL=variable optical attenuator.

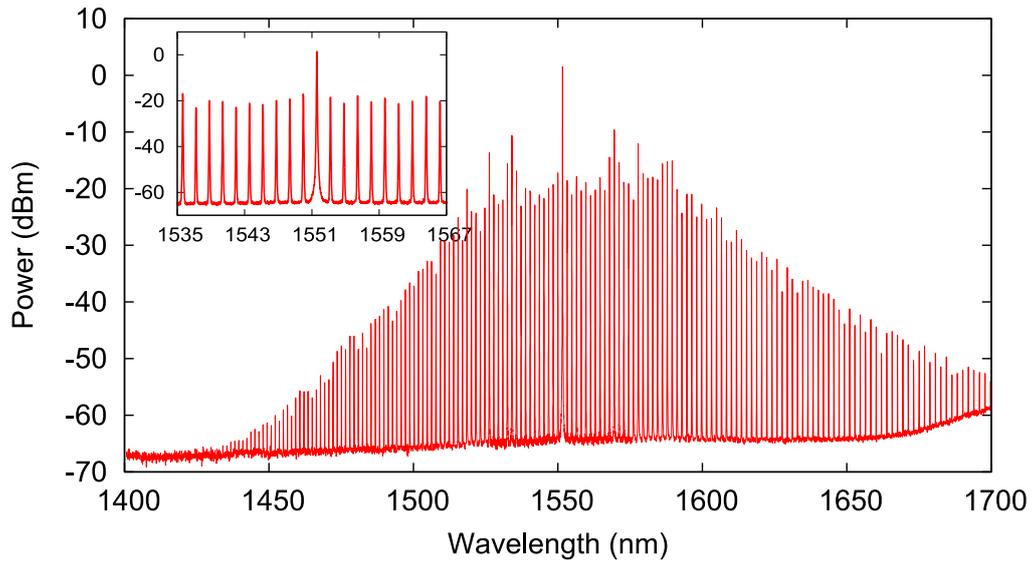

Fig. 4. Spectrum of the comb generated from the microring, measured by an OSA with a resolution of 0.5nm. Inset: Zoom-in of the spectrum around the pump wavelength.

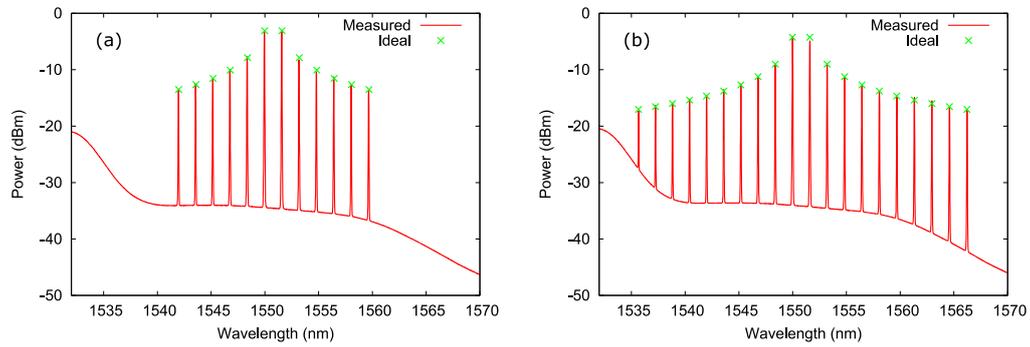

Fig. 5. EDFA2 output showing the weight of each tap for: (a) 12 tap filter, and (b) 20 tap filter

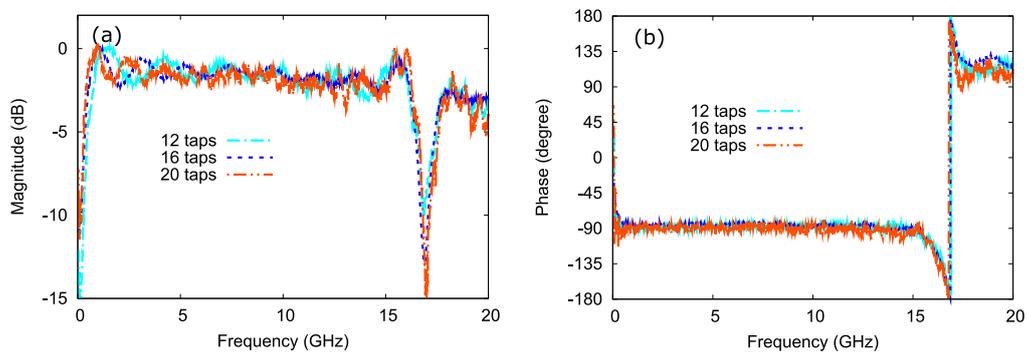

Fig. 6. Measured system RF frequency response for different number of filter taps: (a) amplitude; and (b) phase response

The optical signals were finally detected by photodiodes to regenerate output RF signals. The system RF frequency response was then measured with an RF vector network analyzer (VNA). The maximum power at the drop-port output at resonance was measured to be approximately 14 dB below the input power. When the pump power was 0.5 W, the spectrum measured at the drop port exhibited a broad frequency comb. Fig. 4 shows the spectrum of the comb, spanning more than 250nm, as measured by an optical spectrum analyzer (OSA, with a resolution of 0.5 nm). To generate the comb, the pump laser wavelength was tuned to one of the resonances near 1550 nm. The inset of Fig. 4 shows a high-resolution scan of the measured comb spectrum around the pump wavelength. The spacing of the comb lines corresponded to the FSR of the microring. The power of the comb line at the pump wavelength was about 20 dB higher than the neighboring lines. The generated comb was a "Type II" comb [67] – where the coherence between the comb lines was not particularly high [68]. However, for the photonic Hilbert transformer demonstrated here, the optical signals from different taps were detected incoherently by the photodiode and so a high degree of coherence between comb lines is not required in order to achieve high RF performance.

## IV. EXPERIMENT

### 4.1. Comb shaping

Since the spectrum of the comb generated by the microring did not follow the hyperbolic function required for the impulse response of a Hilbert transformer [Fig. 2(a)], it was necessary to shape the comb in order to achieve the required tap coefficients. Each individual comb line was selected and shaped to the desired power level by the waveshaper. This could be easily achieved since its resolution (10 GHz) was much smaller than the comb spacing (200 GHz). The normalized power of each comb line, needed to achieve a Hilbert Transform, is given by:

$$p_n = \frac{1}{\pi |n - N/2 + 0.5|} \quad (1)$$

where $N$ is the number of comb lines, or filter taps, used in the filters, and $n = 0, 1, 2, ..., N-1$ is the comb index.

Figures 5(a) and 5(b) show the powers of all the comb lines measured at the output of the EDFA2 using an OSA for the 12 tap and 20 tap filter cases, respectively. The target powers at all wavelengths are also shown in Figs. 5(a) and 5(b) as green crosses. The waveshaper was successfully used to shape the powers of all comb lines to within +/-0.5 dB of the target powers. Unused comb lines were attenuated below the noise floor.

### 4.2. System RF response

Once the comb lines were attenuated in order to provide the correct tap coefficients of the impulse response associated with a Hilbert transform, the system RF frequency response was then characterized. A vector network analyzer (VNA) was used to measure the system RF amplitude and phase frequency response. First, the VNA was calibrated with respect to the reference output $RF_{REF}$ and then the signal output $RF_{SIG}$ was measured with the calibrated VNA.

Figure 6 (a) shows the measured RF amplitude frequency response of the photonic Hilbert transform filters for 12, 16 and 20 taps, respectively, which all exhibit expected behavior. All three filters have the same null frequency at 16.9 GHz, corresponding to the tap spacing of $\Delta t = 1/f_c = 59$ ps. This spacing matches the difference in delay between the comb lines, equal to the ring $FSR = 1.6$ nm, produced by propagation through a 2.122 kilometer long SMF fiber with a dispersion parameter $D = 17.4$ ps/(nm km). The null frequency could be controlled by using a different fiber length to adjust the tap spacing.

All filters show < 3 dB amplitude ripple. As predicted in Fig. 2(c), increasing the number of filter taps increases the filter bandwidth. With a 20 tap filter, the Hilbert transformer exhibited a 3 dB bandwidth extending from 16.4 GHz down to 0.3 GHz, corresponding to more than 5 octaves. It is possible to increase this bandwidth further by using more comb lines in the filter. In our experiment, only a small portion of the generated comb spectrum was actually used to realize the filter taps. The number of filter taps that could be achieved was actually limited by the bandwidth of the waveshaper and the gain bandwidth of the optical amplifier (EDFA2). If desired, it would also be possible to reduce the amplitude ripple within the pass-band by apodizing the tap coefficients from the ideal hyperbolic function [49].

Figure 6(b) shows the measured phase response of filters with different numbers of taps, showing very similar responses. Each shows a relatively constant phase of near -90° within the pass-band. There are some deviations from the ideal -90° phase at frequencies close to zero and particularly for the null frequency $f_c$ = 16.9 GHz. The reasons for the phase errors at the band edges are discussed elsewhere [60].

To assess the stability of the system, the RF response was measured at different times [60]. The RF amplitude and phase frequency responses of the 20 tap filter measured immediately after the system was set up, after 30 minutes and then after 1 hour showed that there was a small variation of up to 1 dB in the RF amplitude frequency response when the system was characterized at these different times, whereas the system shows a similar phase response over the pass band except a small phase variation at the band edges. The small fluctuation in the system response can be attributed mainly to the drift in the bias of the Mach- Zehnder modulator. When the modulator bias drifts the filter tap coefficients will depart from the ideal values resulting in a change in the system RF response. An active bias-controller can in principle be used to minimize the effects of bias drift.

## V. DISCUSSION

We have shown that an integrated optical comb source can be effectively used to provide numerous, high quality optical taps for a microwave photonic transversal filter, thus allowing us to demonstrate a very wide bandwidth RF Hilbert transformer with a 3 dB bandwidth of over 5 octaves from as low as 0.3 GHz to 16.9 GHz. It is extremely difficult to match this performance using electronic or other photonic techniques. In this work, only a small number of the available comb lines from the integrated comb source were utilized to realize the filter taps. This was limited by the finite bandwidth of the configurable filter used to shape the comb spectrum as well as the optical fiber amplifiers. Reducing loss in order to potentially eliminate the amplifier, as well as using a wider bandwidth configurable filter, will both allow more comb lines to be used, resulting in an even broader RF bandwidth. Further improvements to the filter ripple and response near the band edges can be achieved through apodization and compensation of imperfections in the modulation and transmission system. Since the integrated comb source can generate many more comb lines than the number of filter taps, it is possible to realize multiple parallel filters using only a single comb source, further reducing the device complexity.

Although our device is still relatively bulky due to the discrete components that were employed, such as the waveshaper, the integrated nature of the comb source has significant potential to reduce the system complexity by combining many different functions on an integrated chip. In addition to enabling high quality comb sources, the high nonlinearity of the Hydex platform as well as that of other integrated comb source platforms is also attractive for additional on-chip signal processing functions. For example, four wave mixing in highly nonlinear waveguides and ring resonators can be combined with a Hilbert transformer to realize devices capable of instantaneous frequency measurements [51].

## VI. CONCLUSION

We review promising RF and microwave applications for CMOS compatible microresonator based frequency combs based on the Hydex glass platform. These devices have significant potential for applications requiring CMOS compatibility for both telecommunications and on-chip WDM optical interconnects for computing and many other applications.